\documentclass{article}
\usepackage{spconf,amsmath,graphicx}
\usepackage{subfigure}
\usepackage{graphicx}
\usepackage{multirow}
\usepackage{booktabs}

\title{U2-KWS: Unified Two-pass Open-vocabulary Keyword Spotting 
\\ with Keyword Bias}
\name{
\begin{tabular}{c}
  \it Ao Zhang$^1$, Pan Zhou$^2$, Kaixun Huang$^1$, Yong Zou$^2$,Ming Liu$^2$, Lei Xie$^{1*}$\thanks{$^*$: Corresponding author.}
\end{tabular}
}

\address{
  $^1$Audio, Speech and Language Processing Group (ASLP@NPU), School of Computer Science, \\Northwestern Polytechnical University, Xian, China\\
  $^2$Space AI, Li Auto
}
\copyrightnotice{979-8-3503-0689-7/23/\$31.00~\copyright2023 IEEE}
\begin{document}
%
\maketitle
\begin{abstract}
Open-vocabulary keyword spotting (KWS), which allows users to customize keywords, has attracted increasingly more interest. However, existing methods based on acoustic models and post-processing train the acoustic model with ASR training criteria to model all phonemes, making the acoustic model under-optimized for the KWS task. To solve this problem, we propose a novel unified two-pass open-vocabulary KWS (U2-KWS) framework inspired by the two-pass ASR model U2. Specifically, we employ the CTC branch as the first stage model to detect potential keyword candidates and the decoder branch as the second stage model to validate candidates. In order to enhance any customized keywords, we redesign the U2 training procedure for U2-KWS and add keyword information by audio and text cross-attention into both branches. We perform experiments on our internal dataset and Aishell-1. The results show that U2-KWS can achieve a significant relative wake-up rate improvement of 41\% compared to the traditional customized KWS systems when the false alarm rate is fixed to 0.5 times per hour.

\end{abstract}
\section{Introduction}
\label{sec:intro}

Keyword spotting (KWS) is the task of detecting predefined keywords from a consecutive audio stream, which is widely applied in edge devices with a speech interface. Traditional KWS approaches are based on keyword/filler models which involve modeling keyword and non-keyword segments with hidden Markov model (HMM)~\cite{hmm1,hmm2}. With the advance of deep learning, many works are proposed to build systems based on a single neural network without HMM and directly predict the keyword or sub-word tokens of the keyword~\cite{0Convolutional,2014Small,wei2021end,jose2020accurate,berg2021keyword}. Nowadays, the majority of keyword spotting systems primarily rely on such deep learning approaches. While these systems can achieve high precision on the pre-defined keywords, they require a large amount of specific training data that contains those pre-defined keywords. Additionally, keywords cannot be changed after training, which requires the user to remember specific keywords. To deal with such limitations, \textit{open-vocabulary} keyword spotting, which allows users to customize keywords, has gained popularity in recent years.


Many previous works for open-vocabulary KWS employ query-by-example (QbyE) methods, which use only audio signals as input~\cite{qbe1,qbe2}. QbyE methods enroll reference keyword speech and compare it with new input speech queries. These methods require users to record the speech of keywords during customization and their performance is influenced apparently by the reference keyword speech. In contrast, customizing keywords based on text makes it easier to set keywords and is stable for different speakers.
A common text-based open-vocabulary KWS method usually adopts an acoustic model from ASR to transform speech into phonetic posteriorgrams and then leverage a post-processing technique such as HMM to predict keyword existence~\cite{lstm-kws,am1,am2,am3}. In practice, the system with a single streaming acoustic model produces a large number of false alarms~\cite{liu2021rnn,tian2021improving}.
Therefore, the multi-stage strategy has been previously adopted to reduce false alarms~\cite{liu2021rnn,2020Multi, yang2021keyword,sigtia2021progressive}. In general, the first stage is a lightweight always-on keyword detector. Once a keyword candidate is detected, the corresponding audio segment is sent to the following stage(s) for further verification.
Multi-stage systems have several modules and complex structures, making them hard to build. Recently, there has been increasing interest in unifying multi-stage modules into one single model. In this direction, Cascaded Transducer-Transformer (CATT-KWS) uses two-pass models, which unify streaming and non-streaming ASR approaches~\cite{U2,unify2}, to unify multi-stage KWS into one model~\cite{catt}. Specifically, it uses the streaming part, which is originally used to generate streaming hypotheses, as the first-stage model to detect possible keywords, and then uses the non-streaming parts, which are originally used to re-score streaming hypotheses, as the validation stages for further verification of keywords detected in the first stage. Although this approach changes the inference of two-pass models from the original recognition \& rescoring to the multi-stage KWS, the two-pass model is still trained according to the criterion of ASR, and the model does not explicitly utilize the information of the keywords. The model without knowledge of keywords is optimized for the accuracy of all words, thus resulting in the under-optimization problem for the KWS task.

Therefore, it is important to particularly integrate the keyword information into the acoustic model to make it more sensitive to the user-defined keywords. Recent works have employed cross-attention techniques~\cite{attention} to incorporate keyword representation in the context of acoustic representation. These approaches can be categorized into two categories based on the type of query employed for cross-attention: acoustic-query attention and keyword-query attention.
The acoustic-query attention employs the acoustic representation as the query, utilizing the keyword information to bias the acoustic model towards the user-defined keywords~\cite{he2017streaming,liu2021rnn,tian2021improving,sjbias}. This attention method can process audio frame by frame, and the output of the attention mechanism is used to predict phonetic posteriorgrams. Because the information in one frame is relatively limited, such attention cannot adequately model the relationship between the entire audio and the keyword.
In contrast, the keyword-query attention employs the keyword representation as the query and performs cross-attention with the entire audio. This method effectively models the dependencies between keywords and audio and the attention output represents the temporal correlation patterns between acoustic representations and keywords, which can be used for direct keyword prediction~\cite{e2e2,e2e, agree, dynamic}. This approach queries entire audio and thus better models global relationships. However, due to its global attention mechanism, it is not suitable for streaming inference.

In this paper, we propose the unified two-pass open-vocabulary keyword spotting framework (U2-KWS). Unlike the previously mentioned multi-stage KWS model, our approach focuses on improving the two-pass model specifically for the KWS task rather than relying on an acoustic model trained with the criterion of ASR. 
First, we employ attention-based keyword bias methods to bias the model towards user-defined keywords. Unlike previous works that use single-type attention, we apply the acoustic-query attention in the streaming branch and the keyword-query attention in the decoder branch within our framework. This allows us to leverage the strengths of each method, considering both streaming capability and accuracy.
Besides, we redesign the training and inference of the two-pass model.  We only feed the attention decoder with the keywords and corresponding acoustic representations, thus training it as a specific keyword rescoring module focusing on the task of keyword verification. We simulate the process of customizing keywords by sampling keywords from transcripts to generate positive and negative samples during training.
We evaluate the proposed U2-KWS framework on the AISHELL-1 dataset~\cite{aishell1} and an internal in-car speech communication dataset. Results on the proposed approach demonstrate an impressively high wake-up rate and low false alarm rate.

\section{U2-KWS}
\label{sec:method}
In this section, we provide a detailed introduction to our proposed U2-KWS framework. First, we provide an overview of the entire architecture of the two-pass keyword spotting framework. Next, we discuss the use of keyword-bias methods in both streaming and non-streaming branches. Finally, we present the newly designed training and inference procedures for the two-pass keyword spotting.

\subsection{Model Architecture}
\begin{figure}[tb]
    \centering
    \includegraphics[width=0.8\linewidth]{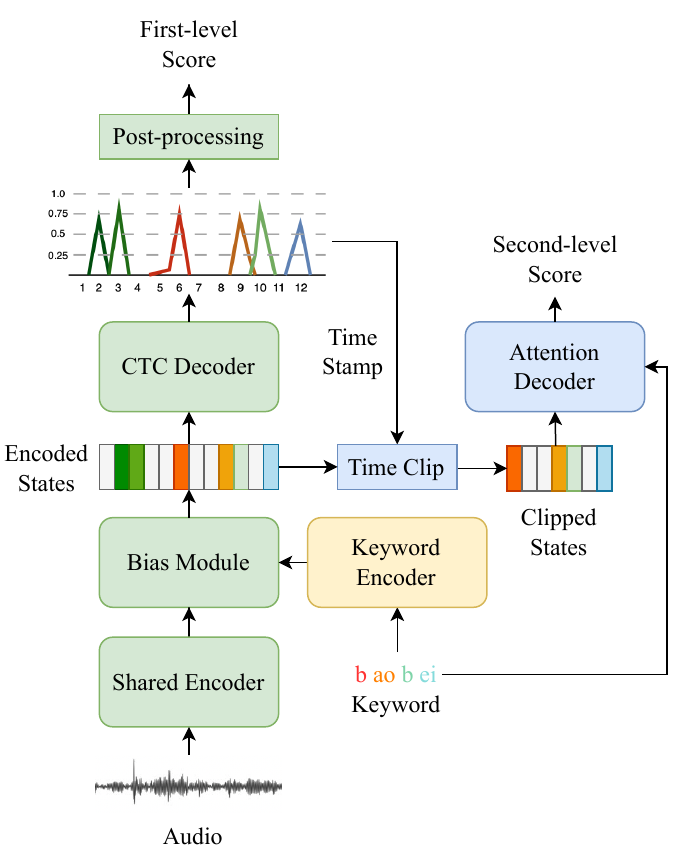}
    \vspace{-10pt}
    \caption{Block diagram of the proposed U2-KWS.}
    \vspace{-5pt}
    \label{fig:overview}
    \vspace{-10pt}
\end{figure}
The proposed model architecture is shown in Fig.~\ref{fig:overview}. It contains four parts: a shared encoder, a bias module, a CTC decoder, and an attention decoder. The green part of the figure is the streaming CTC branch in the two-pass model, which is employed as the first stage model for detecting potential keyword candidates in the audio, while the blue part is the non-streaming branch in the two-pass model, which works as the second-stage model by verifying the detected candidates with an attention decoder to reduce false alarms. The yellow part represents the keyword encoder which models relationships within keywords and encodes keywords into high-level representations.

The shared encoder consists of multiple Conformer~\cite{conformer} layers and is trained with the dynamic chunk strategy~\cite{wenet} which enables latency control and can forward in different chunk sizes. The keyword encoder is a single LSTM. The bias module consists of a multi-head attention for model bias and a linear layer for dimensional transformation. The CTC decoder consists of a linear layer and a log-softmax layer, which outputs the phonetic posteriorgrams. The attention decoder is a Transformer decoder~\cite{attention} but gets keywords instead of the hypotheses as input and does cross-attention with acoustic representations corresponding to the keyword phrase.
\subsection{Customized Keyword Bias}
In this section, we will provide a detailed explanation of the keyword bias methods based on acoustic-query attention and keyword-query attention. Afterward, we will discuss how we unify these methods into our two-pass model.
\subsubsection{Attention-based Keyword Bias}
The cross-attention mechanism is a widely used and efficient approach for modeling inter-modal relationships. In this paper, we formalize the cross-attention mechanism as:
\vspace{-10pt}
\begin{equation}
\operatorname{Attention}(Q, K, V)=\operatorname{softmax}\left(\frac{Q\ K^T}{\sqrt{d_k}}\right)\ V.
\end{equation}
Here, ${Q}$ represents the query vector, ${K}$ represents the key vector, and $V$ represents the value vector. We will now discuss the attention-based keyword bias method.

The acoustic-query attention method employs the acoustic representation ${h_a}$ as the query and the keyword representation ${h_k}$ as the key and value, which can be formulated as:
\begin{equation}
{h_a}\xrightarrow{\text{linear}}Q_a;h_k\xrightarrow{\text{linear}}K_k,V_k,
\end{equation}
\begin{equation}
    \tilde{h_a}=\text{Attention}(Q_a,K_k,V_k).
\end{equation}
The resulting attention output ${\tilde{h_a}}$ retains the same dimension as the acoustic representation and can be combined with ${h_a}$ for the prediction of phonetic posteriorgrams.

The keyword-query attention method utilizes the keyword representation ${h_k}$ as the query and the acoustic representation ${h_a}$ as the key and value. The resulting attention output ${\tilde{h_a}}$ represents the agreement between the audio and text~\cite{agree}. 
This process can be described as follows:
\vspace{-5pt}
\begin{equation}
{h_k}\xrightarrow{\text{linear}}Q_k;h_a\xrightarrow{\text{linear}}K_a,V_a,
\end{equation}
\begin{equation}
    \tilde{h_k}=\text{Attention}(Q_k,K_a,V_a).
\end{equation}
\vspace{-20pt}
\subsubsection{Two-pass Keyword Bias}
\begin{figure}[tb]
    \centering
    \includegraphics[width=1.0\linewidth]{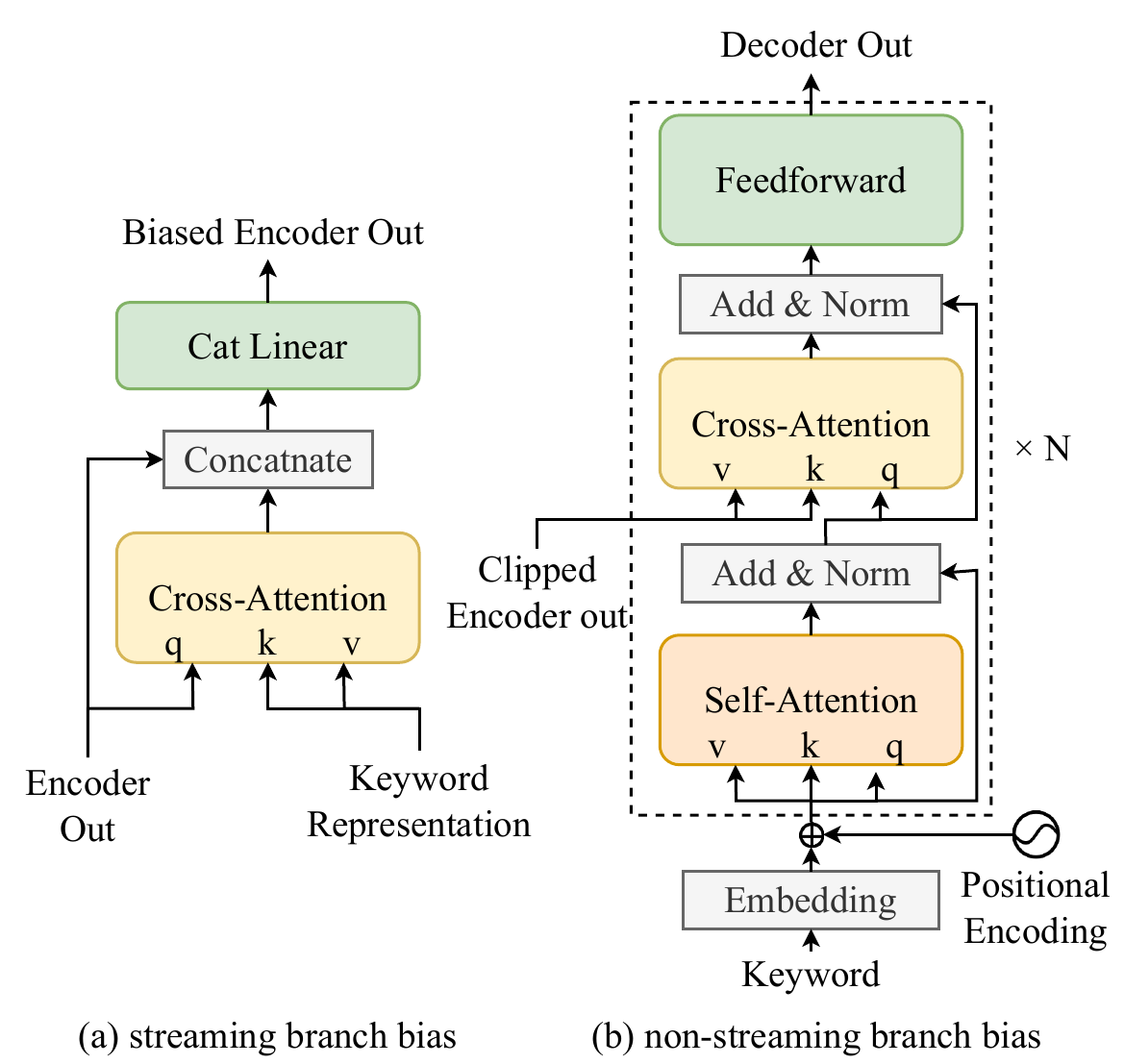}
    \vspace{-15pt}
    \caption{Block diagram of keyword bias in U2-KWS.}
    \vspace{-5pt}
    \label{fig:multi-att}
    \vspace{-5pt}
\end{figure}
The applications of acoustic-query attention and keyword-query attention in our framework are shown in Fig.~\ref{fig:multi-att}. To incorporate the acoustic-query attention, we integrate it into the streaming branch of the two-pass model. Firstly, we encode the customized keyword into an embedding using the keyword encoder. Next, we utilize the acoustic-query attention in the bias module to integrate the keyword information into the acoustic representation encoded by the shared encoder. The output of the attention module is concatenated with the original acoustic representation and then projected back to the original dimension. The resulting biased encoder output includes both the acoustic and keyword information and is used for subsequent inference.

For the keyword-query attention, we use it in the modified transformer decoder to bias the non-streaming branch. We keep the original transformer decoder structure and fix its input as keywords instead of the decoding results of the steaming branch, thus making the decoder a specialized keyword re-scoring module. In this case, the self-attention module encodes keywords into high-level representations and the cross-attention is the keyword-query attention. The output of the decoder is the token-level posterior probability, and we calculate the possibility of the keyword path as the score of the second stage model. This approach more adequately models the correlation between keywords and speech but is also more resource-intensive so we only used this structure for the non-streaming branch.

\subsection{Training and Inference}
\subsubsection{Training of First Stage}
We train the streaming branch first. The streaming branch is optimized with CTC loss, and since we introduced the keyword-bias module we need to generate keyword samples during training. To generate positive and negative samples, we employ a sampling technique on the transcripts to simulate user-defined keywords. Specifically, for a given utterance, we randomly select a consecutive word sub-sequence from its transcript as the keyword to create a positive sample. For generating a negative sample, we randomly combine words from the lexicon that are not present in the transcribed text of the utterance. To introduce a training error for keyword spotting and encourage the model to pay attention to the keyword information, we add a \textless eok\textgreater\enspace token after the sampled keyword, which is similar to the label-augment method proposed for context bias in CLAS~\cite{clas}.
To provide a clearer understanding of this process, we take the sample ``Call you Jarvis" as an example. The positive and negative samples produced for this sample are illustrated in Table~\ref{tab:sample}. To be concise, we employ words as an instance for illustrating purposes, as we actually convert words into phonemes during training and inference.\vspace{-10pt}
\begin{table}[h]
\caption{Example of keyword sampling}
\renewcommand{\arraystretch}{1.5}
\begin{tabular}{lcc}
\toprule
Class                                                           & Positive                                    & Negative                                                 \\ \hline
Text                                                            & \multicolumn{2}{c}{Call you Jarvis}                                                                    \\ \hline
Keyword                                                         & Jarivs                                      & Alex                                                     \\ \hline
\begin{tabular}[l]{@{}l@{}}Keyword\\ Encoder Input\end{tabular} & Jarvis\textless{}eok\textgreater{}          & Alex\textless{}eok\textgreater{}                         \\ \hline
CTC Target                                                      & Call you Jarvis\textless{}eok\textgreater{} & Call you Jarvis                                          \\ \hline                  
Decoder Input                                                   & \textless{}sos\textgreater Jarvis           & \textless{}sos\textgreater{}Alex                         \\ \hline
Decoder Target                                                  & Jarvis\textless{}eok\textgreater{}          & \textless{}eos\textgreater{}\textless{}eos\textgreater{} \\ \bottomrule
\end{tabular}
\vspace{-18pt}
\label{tab:sample}
\end{table}
\subsubsection{Training of Second Stage}
\label{sec:decoder_training}
After the training of the streaming branch, we jointly optimize the whole two-pass model with attention loss and CTC loss. In the training phase, we follow the same method to generate positive and negative samples as before. The overall forward and loss calculations of decoder training remain the same with the attention decoder in the ASR task, but we make modifications to the input token and target. Specifically, we configure the input of the decoder to be a fixed sequence consisting of ``\textless sos\textgreater + keyword". For positive samples, the target is set as ``keyword + \textless eok\textgreater", while for negative samples, the model directly predicts \textless eos\textgreater, as shown in Table~\ref{tab:sample}. In this way, the decoder focuses on the prediction of keyword sequence during training and becomes a token-level classifier with keyword discrimination capability.

Inspired by the end-to-end segmentation method used in the two-pass ASR model~\cite{seg}, we use timestamp information from the streaming branch to clip the encoder output for the decoder branch. This allows the cross attention of the decoder to only focus on the acoustic representation that contains the keyword. By doing so, we can reduce the complexity of the task for the decoder and improve keyword spotting performance.

To achieve the clip of the encoder output, we require the start and end times of the keyword. With the \textless eok\textgreater\enspace token in the first-level model, we can identify the frame with the highest posterior probability of \textless eok\textgreater\enspace as the end frame. To estimate the start frame, we use a method similar to the spike trigger CTC~\cite{spike}, where we consider a spike as any non-blank token with a posterior probability exceeding a pre-set threshold. We count the number of spikes from the end frame back to the start of the speech, and when the number of spikes exceeds the length of the keyword sequence, we consider the last counted spike as the starting frame. This method doesn't need extra alignment information and can output stable lengths compared with the alignment method based on CTC posterior probability.

The joint loss function for the entire two-pass model can be expressed as:

\begin{equation}
\setlength\abovedisplayskip{6pt}
\setlength\belowdisplayskip{6pt}
L=\lambda L_{ctc}+(1-\lambda) L_{att},
\end{equation}
where $L_{ctc}$  and $L_{att}$ represent the CTC loss and the attention loss, respectively. The hyper-parameter $\lambda$ controls the contribution of each loss to the overall training objective.
\vspace{-10pt}
\subsubsection{Inference}
\label{sec:infer}
In the initialization phase, the keyword encoder generates a high-level representation of the customized keyword entered by the user, which is marked in yellow in Figure~\ref{fig:overview}. The first stage model, highlighted in green in the same figure, processes the input audio stream chunk by chunk to get the posterior probability of CTC. We search the keyword path on the post probability to get the path with the highest probability and use the probability as the score of the first stage. If the score surpasses the threshold set for the first stage, we clip the acoustic representation based on the \textless eok\textgreater\enspace token and non-blank token spikes of CTC. Then, we employ the decoder to conduct keyword rescoring by sequentially feeding the keyword sequence token into the decoder and computing the probability of the next token of the keyword. This score is compared to the threshold for the final prediction. Since our encoder is trained with dynamic chunks, the encoder can infer in full-chunk mode to fully utilize the context. We can re-feed audio segments containing keyword candidates detected in the first stage to the encoder again in full-chunk mode to get more informative acoustic representations for the rescoring of the decoder in the second stage. Thus we further improve performance by only introducing minimal latency. We call the direct use of streaming encoder output the causal decoder, and the re-feeding method the full decoder.

\vspace{-8pt}
\section{Experiments}
\label{sec:exp}
In this section, we introduce the corpus and describe the experimental setup including our model configuration. Experimental results and analysis are also presented at last.
\subsection{Corpus}
\textbf{Internal Corpus}:  Models are trained on a Mandarin ASR corpus comprising 1,000 hours of speech collected in hybrid electric vehicles. We randomly shuffled the development set from the training set.
To evaluate the accuracy of our models, we record the positive test set in the same environment as the training set. This test set contains 30 different keywords, each including 200 samples, for a total of 3,000 positive samples. These keywords are composed of two to four Chinese characters.
To evaluate false alarms, we used a separate audio set of 60 hours, which mainly contained chat, command, and radio broadcasts, as the negative test set.
All the data mentioned above, including the training set, development set, positive test set, and negative test set, are anonymous and hand-transcribed.

\textbf{AISHELL-1}: We also conduct experiments on the public Chinese Mandarin speech corpus AISHELL-1~\cite{aishell1}. For training and validation, we use the train set and dev set of AISHELL-1. For evaluation, we construct a test set consisting of 7,176 pairs of positive and negative samples from the AISHELL-1 test set using the keyword sampling method adopted in training.
\subsection{Configuration}
Our model used 80-dimensional log Mel-filter banks with a 25ms window and a 10ms shift. SpecAugment~\cite{specaug} is applied 2 frequency masks with maximum frequency mask (F = 10), and 2 time masks with maximum time mask (T = 50) to alleviate over-fitting. Two convolution sub-sampling layers with kernel size 3*3 and stride 2 are used in the front of the encoder. 

The baseline CTC model we used consists of a 12-layer conformer with 128-dimensional input, 4 self-attention heads, and 256 linear units. The larger baseline expands the input dim to 256 and the number of linear units to 2048 which makes it nine times larger than the baseline model. For the keyword bias, the keyword encoder comprises an embedding layer and one layer LSTM with a dimensionality of 128. The bias module is a multi-head attention layer of 128 hidden states and 4 attention heads. For the decoder branch, we employ 2 transformer layers with 512 linear units as the decoder. All models are trained with dynamic chunks and conducted inference in chunks of 8. 

The output units include 210 context-independent (CI) phones, a \textless sos/eos\textgreater symbol, and a \textless eok\textgreater symbol.
The model consists of multiple components with their respective parameter sizes. The encoder has a size of 3.75M, the CTC decoder has a size of 0.2M, the attention decoder has a size of 0.6M, the keyword encoder has a size of 0.16M, and the bias module has a size of 0.04 M. In total, the model has 4.75M parameters. 
It is important to note that the keyword encoder only works for the system initialization, and the decoder is activated only when the streaming branch detects keyword candidates. This allows the two-pass model to have only 0.04M extra always-on parameters compared to the baseline model, making it suitable for deployment on edge devices.
\subsection{Evaluation Metrics}
We evaluate the performance in terms of the receiver operating characteristic (ROC) curve and F1-score. To generate the ROC curves for each keyword, we utilize corresponding positive samples and the 60-hour negative test set. To evaluate the overall performance of different keywords, we average the false rejection rates of all keywords at the same false alarm rate. This allows us to plot overall ROC curves, providing a comprehensive evaluation of models.
Similar to the process of ROC curves, we compute the F1-score for each keyword and average them to evaluate the overall performance.
\subsection{Performance of Two-pass KWS Framwork}

\begin{figure}[t]
    \centering
    \includegraphics[width=0.9\linewidth]{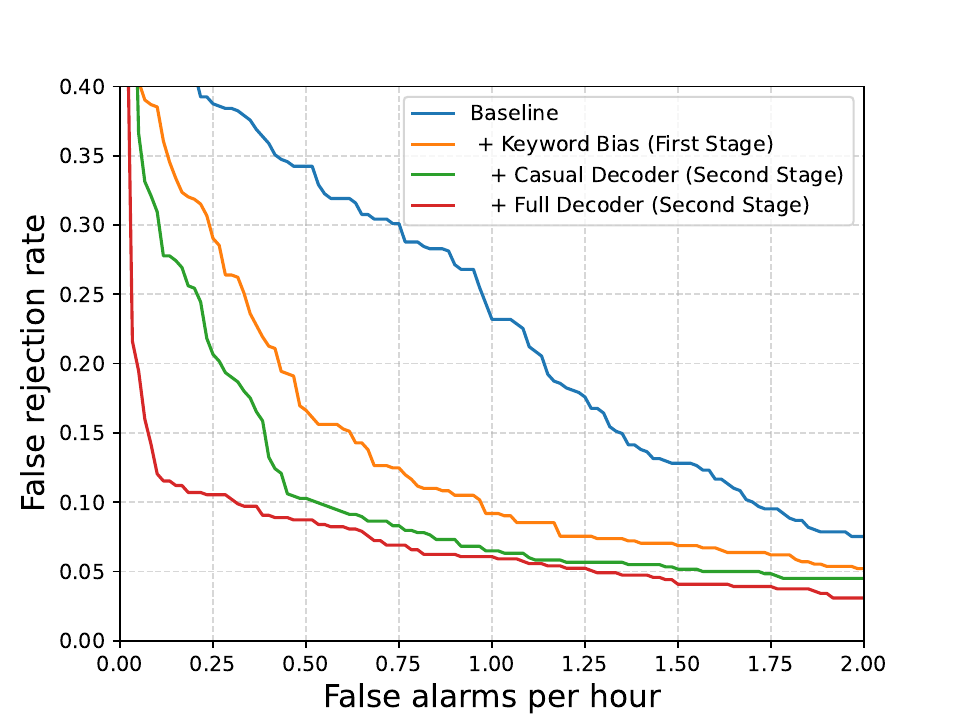}
    \caption{ROC curves for different systems.}
    \vspace{-5pt}
    \label{fig:over1}
\end{figure}

In Fig.~\ref{fig:over1} and Table~\ref{tab:len}, we show the results of our proposed framework and baseline model on the internal test sets. The baseline system consists of an acoustic encoder with the CTC decoder. Integrating keyword information with keyword bias into the baseline model we get the first stage model. Compared with the baseline
model, the first stage model increases the wake-up rate by 20\% under the condition of 0.5 false alarms per hour. This shows that integrating keyword information through attention is a very effective method for open-vocabulary keyword spotting. 
Both the causal decoder and the full decoder introduced in Sec~\ref{sec:infer} can further improve performance on the first stage model. The causal decoder utilizes the output of the streaming encoder for rescoring, making better use of contextual information and achieving a 6\% improvement in wake-up rate under a 0.5 false wake-up condition. On the other hand, the full decoder uses the full-chunk encoder to obtain more accurate acoustic representations, resulting in a 10\% improvement under a 0.25 false wake-up condition.

\subsection{Impact of Varying Keyword Length}
To investigate the impact of keyword lengths on system performance, we conducted experiments by grouping the test sets based on the character number of keywords and evaluated them separately. In Table~\ref{tab:len}, we present the average F1-scores for each keyword length.
\begin{table}[t]
\caption{F1-scores of different systems on varying lengths}
\begin{tabular}{lcccc}
\toprule
\multirow{2}{*}{Model} & \multicolumn{4}{c}{F1-score in different word length} \\ 
                       & Overall    & 2        & 3        & 4   \\ \midrule
Baseline               &   0.790    & 0.711    &0.813        &0.847     \\
\hspace{0.5em}+ Encoder Integration   &   0.872    &  0.791   &0.893     & 0.932    \\
\hspace{1em}+ Casual Decoder       &   0.894    &  0.793   &0.926     & 0.963    \\
\hspace{1em}+ Full Decoder         &   0.910    &  0.807   &0.946     & 0.977    \\ \bottomrule
\end{tabular}
\label{tab:len}
\end{table}
Our findings indicate that longer keywords tend to result in better overall performance. This can be attributed to the fact that shorter keywords are more prone to generating false alarms. We can observe that the longer the keywords are, the greater the improvement the decoder branch brings.
There are two reasons for this. First, for short sequences, the structural benefit of the decoder which makes full use of context becomes smaller, and the second is that long sequences need to calculate more steps in the decoder, which can give full play to the distinguishing ability of the decoder.

\subsection{Impact of Decoder Strategy}
\begin{figure}[t]
    \centering
    \includegraphics[width=0.9\linewidth]{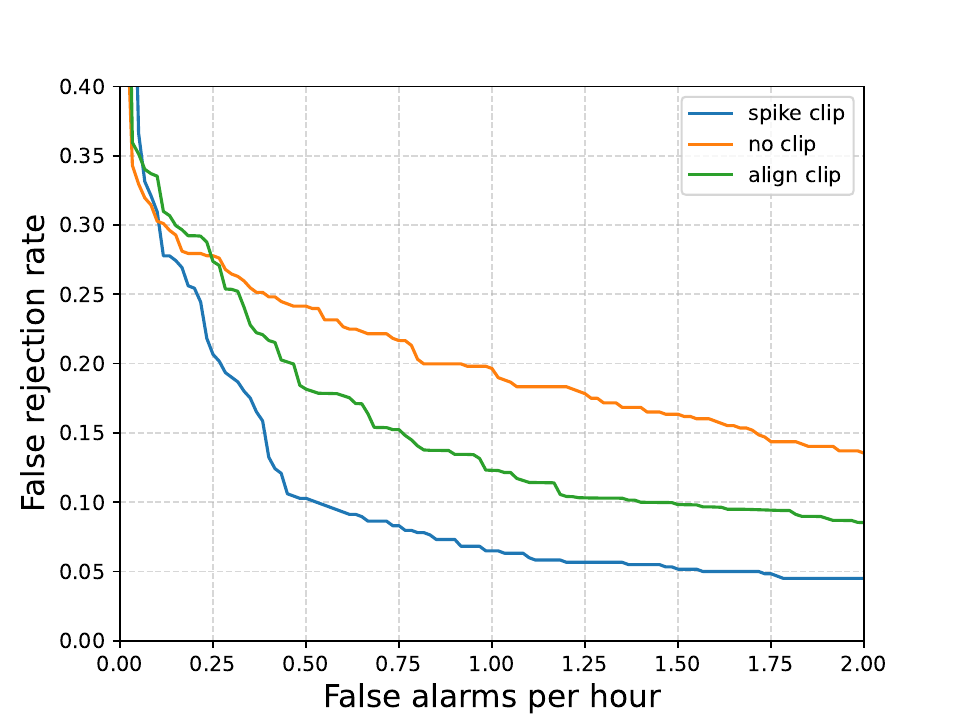}
    \caption{ROC curves for systems with different clip strategies.}
    \vspace{-5pt}
    \label{fig:jieduan}
    \vspace{-5pt}
\end{figure}

We conduct comparative experiments with the causal decoder to explore various methods of clipping the encoder output during decoder training and inference. The results of these experiments are depicted in Fig~\ref{fig:jieduan}.
The method without clipping assigns the task of searching keyword-related acoustic representations to the cross-attention of the decoder, which increases the complexity of decoder tasks. Using timestamp information in the first stage to clip the output of the encoder can make the decoder focus on the keyword validation task, so the clip method is better than the non-clip method. Although the method based on CTC alignment can get accurate timestamps for clean positive samples, it will output too long or too short clipping ranges for negative samples and difficult positive samples, resulting in decoder instability. In contrast, the spike-based methods provide a more stable clipping range by deriving the timestamp based on the number of peaks rather than specific content. Although this method may clip redundant parts, the cross-attention of the decoder is resilient to a small amount of redundant information. Thus, the spike-based method yields the best results.

\subsection{Performance on AISHELL-1}
\begin{figure}[t]
    \centering
    \includegraphics[width=0.9\linewidth]{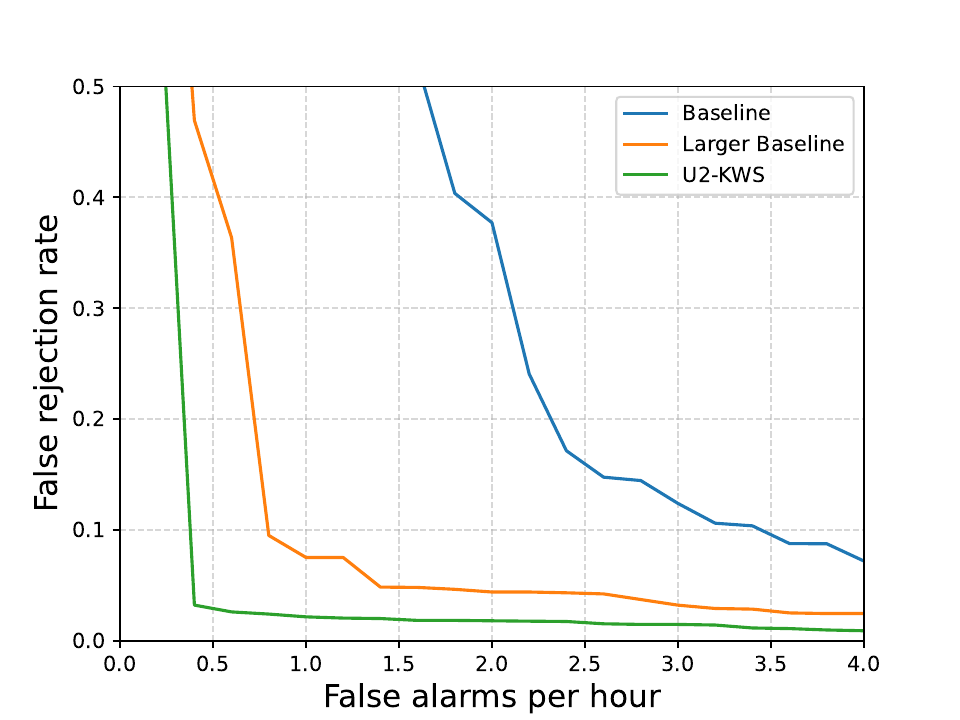}
    \caption{ROC curves for different systems on AISHELL-1.}
    \vspace{-5pt}
    \label{fig:aishell}
    \vspace{-10pt}
\end{figure}

To ensure our findings are reproducible, we carry out additional experiments using the publicly available Mandarin speech corpus AISHELL-1. The ROC curve presented in Fig.~\ref{fig:aishell} clearly demonstrates that our proposed framework U2-KWS significantly outperforms the baseline. These results are consistent with the conclusions drawn from our analysis of the internal dataset.

\section{conclusiosn}
\label{sec:conclusion}

In this paper, we propose a novel two-pass open-vocabulary keyword spotting framework U2-KWS. The framework uses the streaming branch as the first stage model to detect keyword candidates, while the non-streaming branch is activated to make further validation only when the first stage model has detected keyword candidates. To make the model sensitive to the keywords, we introduce keyword bias into both branches and design a two-pass training process for KWS with keyword sampling and encoder clipping. Our method outperforms the baseline by a significant margin on both internal and open datasets.  We also set up additional experiments to explore the effectiveness of the decoder strategy and the impact of different keyword lengths.

\bibliographystyle{IEEEbib}
\bibliography{strings,refs}

\end{document}